\documentclass{article}
\usepackage[utf8]{inputenc}
\usepackage{graphicx}
\usepackage{tabularx}
\usepackage{stringenc}
\usepackage[filenameencoding=utf8]{grffile}
\usepackage{listings}
\usepackage{float}
\usepackage{hyperref}

\newcommand{\function}[1]{\textit{\sc{#1}}}
\newcommand{\service}[1]{#1}
\newcommand{\repo}[1]{\texttt{#1}}
\newcommand{\filename}[1]{\texttt{#1}}
\newcommand{\syntax}[1]{'\texttt{#1}'}

\newcommand{\liveupdate}[0]{\textit{live update available}}
\hypersetup{
pdfborder={0 0 0.5 [3 3]},
urlbordercolor={0.7 0.7 0.7},
citebordercolor={0.9 0.9 0.9},
linkbordercolor={0.9 0.9 0.9}
}

\title{Quantitative Analysis of Cloud Function Evolution in the AWS Serverless Application Repository}
\author{Josef Spillner\\
Zurich University of Applied Sciences,\\
Service Prototyping Lab (\href{https://blog.zhaw.ch/splab/}{blog.zhaw.ch/splab}), Switzerland\\
\href{mailto:josef.spillner@zhaw.ch}{josef.spillner@zhaw.ch}}

\begin{document}

\maketitle

\begin{abstract}
The serverless computing ecosystem is growing due to interest by software engineers. Beside Function-as-a-Service (FaaS)
and Backend-as-a-Service (BaaS) systems, developer-oriented tools such as deployment and debugging frameworks
as well as cloud function repositories enable the rapid creation of wholly or partially serverless applications.
This study presents first insights into how cloud functions (\service{Lambda} functions) and composite serverless applications
offered through the \service{AWS Serverless Application Repository}
have evolved over the course of one year. Specifically, it outlines information on cloud function and function-based application offering models
and descriptions, high-level implementation statistics, and evolution including change patterns over time.
Several results are presented in live paper style, offering hyperlinks to continuously updated figures to follow
the evolution after publication date.
\end{abstract}

\section{Introduction}

Software and service marketplaces are increasingly used to rapidly assemble powerful online applications.
Initially, the focus has been on downloadable software artefacts only \cite{DBLP:conf/codaspy/ShuGE17,DBLP:journals/corr/abs-1903-05394},
but increasingly, developers expect
brokered software running as hosted and managed service instances with modest configuration effort \cite{DBLP:journals/dke/BassiliadesSGKM18}.
Owing to the popularity of serverless computing \cite{faascomparison}, marketplaces for cloud functions are particularly
suited to accommodate this need \cite{faasecosystems}. Such functions are considered small, re-usable and composable entities
whose ephemerality and statelessness are attractive to developers of extensible applications, event-driven systems
and scientific workflows \cite{DBLP:journals/cse/Vazquez-Poletti18}.
Although in practice cloud functions have issues such as lack of portability
and arbitrary resource limits in the Function-as-a-Service (FaaS) hosting environments of many commercial public clouds,
developers recognise the potential for improvement and have come up with workarounds and patterns to many of
them \cite{DBLP:journals/corr/abs-1902-03383}. In conjunction with marketplaces, hubs and repositories offering
readily configured cloud functions as plug-in solution, the FaaS platforms and tools might eventually evolve
into holistic, polyglot and cross-vendor Platform-as-a-Service (PaaS) offerings
closely matching the productivity and simplicity expectations of many developers \cite{DBLP:journals/corr/abs-1803-07680}.

While not the first marketplace for cloud functions \cite{functionhub}, and despite much smaller scale compared to traditional
software artefact repositories such as \service{Maven Central} and \service{Docker Hub}, the \service{Serverless Application Repository} by Amazon Web Services
(AWS SAR\footnote{AWS SAR: \url{https://aws.amazon.com/de/serverless/serverlessrepo/}})
has certainly become the most well-known and widely used representative.
For function developers, SAR allows for public offerings of \service{AWS Lambda} functions,
under the condition of open source implementations. For developers of function-based (serverless) applications,
SAR offers either Lambda-deployable or, for functions marked public, also private FaaS-deployable common dependency functions,
as well as orchestrations involving multiple functions and BaaS subscriptions.
An example of a such a dependency function with high degree of re-usability is the \function{image-moderation-chatbot} function which
removes images with explicit content from chats, so that developers of chat applications can focus on core parts of their applications.
Another example would be a conjunction of retrieving log files to Lambda and sending the processed output to Slack --
a chain of two functions which can be represented as composite application in SAR, as demonstrated by the \function{cw-logs-to-slack} function.

According to AWS, the main advantage of using SAR in general is that it makes several steps superfluous including code cloning,
compilation, packaging and publishing to \service{AWS Lambda}.
Having appeared in early 2018, little is known
about this marketplace and about the cloud functions and function-based applications delivered through it. This absence of documented knowledge
is in contrast to the increasing relevance of integrating cloud functions into software applications \cite{SLD_127,SLD_122},
and furthermore in contrast to an increasingly active community mining software repositories \cite{DBLP:conf/msr/SchermannZC18,DBLP:conf/msr/MaFCAZM08}.

In this quantitative study, the aim is thus to gain insights into how functions are implemented, offered and deployed.
For this purpose, the evolution of function-level metadata, code-level metadata and code-level implementation statistics
of Lambda functions offered through AWS SAR is investigated. For brevity, function-based applications as subsumed under
the term function throughout the study.
Three guiding research questions $RQ_1-RQ_3$ determine the study methods and procedures:

\begin{enumerate}
\item[$RQ_1$] How are cloud functions offered and described on commercial marketplaces? More precisely, which information can be extracted
from metadata and which findings can be derived from a function-level metadata analysis alone?
\item[$RQ_2$] How are cloud functions implemented? More precisely, which programming languages, frameworks and structural code patterns
can be identified?
\item[$RQ_3$] Which change patterns and trends on brokered cloud functions can be recognised over time? More precisely, is the assumed popularity growth of serverless
computing reflected in growing numbers of functions on marketplaces and growing numbers of deployed cloud function services?
\end{enumerate}

To answer these questions, the study first presents the research methods on metrics collection, function-level and code-level metadata analysis and code-level programming analysis. It then presents the results as corresponding answers $A_1-A_3$. The results are then discussed in a broader context to stimulate follow-up work to reveal more detailed knowledge about the implementations. To maintain a high relevance, many result figures link to their continuously updated online counterparts in a live paper style.
In the interest of brevity, no background section on serverless computing or the AWS serverless ecosystem (Lambda, Lambda@Edge, Step Functions, SAR, Serverless Application Model -- SAM, CodePipeline, ...) is included;
peer-reviewed and authoritative background literature is widely available from the regularly curated Serverless Literature Dataset \cite{sldataset}.

\section{Research Method}

The method used is continuous observation in conjunction with extraction, mining and conflation of function repository, code repository and artefact implementation. Hence, this work also contributes to the ongoing global research effort of assessing microservice implementations\footnote{MAO-MAO -- Microservice Artefact Observatory: \url{https://mao-mao-research.github.io/}}.

For the preparation of the quantitative analysis, AWS SAR was observed by the author first manually over a period of almost three months, from its launch at the end of February 2018 to mid-May 2018. Subsequently, the observation has been continued with automated tools over a period of one full year, from mid-May 2018 to mid-May 2019. As the observation is ongoing and the study merely presents a one-year snapshot, it covers the complete evolution timeline of AWS SAR across all offered Lambda functions with an increasing number of metrics and indicators.

\begin{figure}[htb]
\centering
\includegraphics[width=1.0\columnwidth]{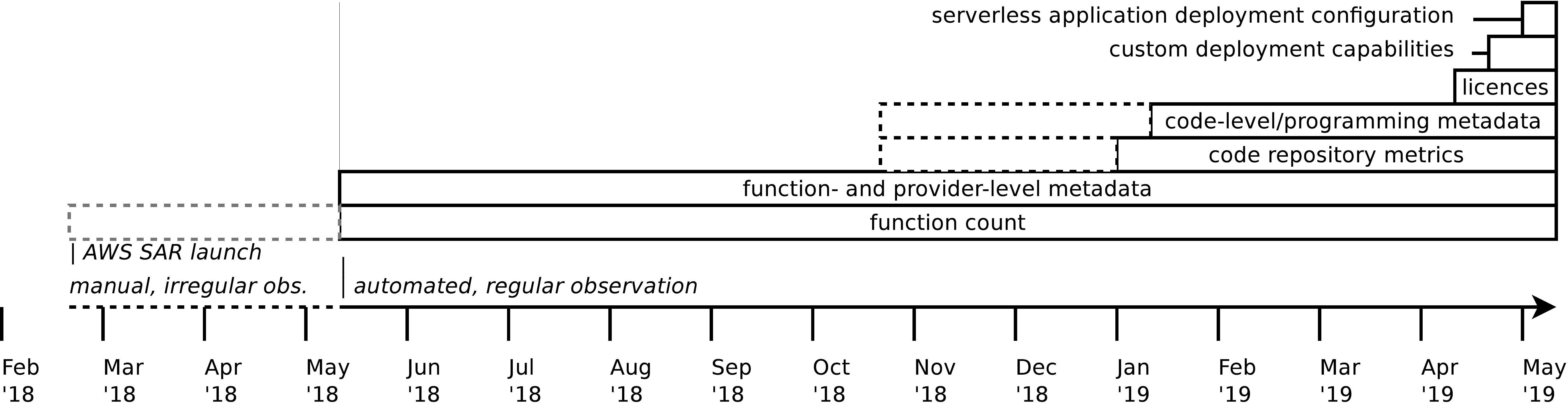}
\caption{Timeline of the experiment on observation and mining AWS SAR\label{fig:timeline}}
\end{figure}

On a daily automated basis, metrics about deployable functions (which do not require custom capabilities)
have been collected, and moreover, function implementation
code repositories publicly available for some of the functions have been tracked and evaluated, in particular on GitHub
which turned out to host the vast majority of function code.
Additionally, from mid-September to mid-December 2018, in-depth dissections of the metadata and code were
initially performed, and from mid-January 2019 on have been an integral part of the automated tracking.
Only since late April 2019, functions with custom capabilities have been added to the automated observation.
Fig. \ref{fig:timeline} shows how the experiment unfolded over many months, including the undesired omission
of historic raw data before the full scope has been achieved.

\begin{figure}[htb]
\centering
\includegraphics[width=1.0\columnwidth]{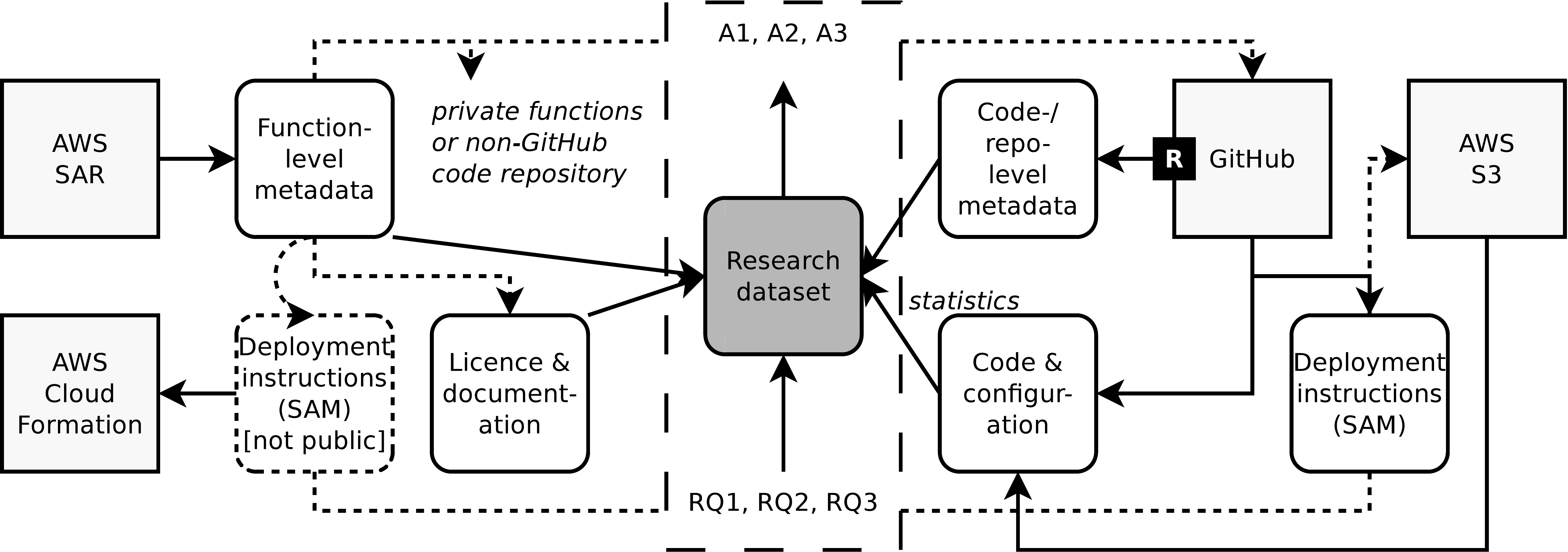}
\caption{Experiment setup to conduct quantitative analysis\label{fig:researchmethod}}
\end{figure}

The full experiment setup at the end of the timeline is shown in Fig. \ref{fig:researchmethod}.
It shows how all research questions $RQ_1-RQ_3$ are answered through a rich dataset which is carried forward
daily to gain up-to-date and increasingly precise insights.
The dataset is produced by the function-level metadata supplied in JSON format by AWS SAR, the referenced licences and README documentation in plain text format, code-level metadata supplied in JSON format by GitHub, and code in various structured formats. Additionally, YAML-formatted deployment instructions are referenced from each function but are only accessible inside the AWS platform. The duration of each daily experiment run is dominated by the rate limitation of the GitHub API, enforcing 61+ second intervals between requests (marked with \textbf{R}). Hence, the duration grows linearly with the number of functions associated to code repositories.

All raw data as well as the aggregation and analysis scripts are available as open research
dataset\footnote{Quantitative analysis research data: \url{https://github.com/serviceprototypinglab/aws-sar-dataset}}\textsuperscript{,}\footnote{Quantitative analysis scripts: \url{https://github.com/serviceprototypinglab/aws-sar-analysis}}.
The author and dataset curator encourages other researchers to perform deeper inspections and to continue the metrics collection for more
targeted studies on both \service{AWS Lambda}-deployable and publicly available (privately deployable) cloud functions.

\subsection{Function metadata collection}

AWS SAR offers a web interface implemented as dynamically generated Scalable Vector Graphics (SVG) user interface
to browse offered functions.
Additionally, a web service query interface with JSON format and enforced pagination is
available as feed endpoint\footnote{AWS SAR feed: \url{https://shr32taah3.execute-api.us-east-1.amazonaws.com/Prod/applications/browse}}.
Upon retrieving this feed regularly, metadata entries with eight metrics become available.
The algorithm for the retrieval of paginated metadata $M$ is given in Listing \ref{lst:retrieval}.

\begin{lstlisting}[caption=Retrieval of AWS SAR feed, label=lst:retrieval, language=python, basicstyle=\ttfamily\small, frame=single, breaklines=true, mathescape=true]
$pages_{needed}$ $\leftarrow$ undef
$entries$ $\leftarrow$ 100 # max 100
$caps$ $\leftarrow$ IAM,NAMED_IAM,RESOURCE_POLICY,AUTO_EXPAND

LOOP $page$ $\leftarrow$ 1..$\infty$ UNTIL $page$ = $pages_{needed}$
	$link$ $\leftarrow$ "https://...FEEDENDPOINT?pageSize=" + $entries$
	if $page$ > 1
		$link$ $\leftarrow$ $link$ + "&pageNumber=" + $page$
	$link$ $\leftarrow$ $link$ + "&includeAppsWithCapabilities=" + $caps$

	$M$ $\leftarrow$ download($link$)

	IF $page$ = 1
		$pages_{needed}$ $\leftarrow$ $\left\lceil{\frac{M_{approximateResultCount}}{100}}\right\rceil$

	LOOP $app$ $\leftarrow$ $M_{applications}$
		$metrics$ $\leftarrow$ $app_{name}, app_{id}, app_{labels}, app_{deploymentCount}, ...$
\end{lstlisting}

The default feed only lists cloud functions which run in unprivileged mode. In contrast, parameterisation
allows for specifying resource capabilities and custom identity and access management rules, including
also functions requiring these at runtime. The algorithm is therefore designed to fetch all functions
while the information about required capabilities is preserved for post-processing.
From the metadata, references to further data sources are extracted and processed, crossing multiple
system boundaries as outlined in the research method figure.

\subsection{Function metadata analysis}

To give an answer to $RQ_1$, the following analysis steps on individual data fields and metrics are carried out on the retrieved metadata $M$:

\begin{enumerate}
\item Distribution of discrete features among $M$, including vendors and their market shares, labels, deployment counts and averaged deployment ratios
\item Short description text metrics such as length and language
\item Referenced long description text and licence text metrics, including a distribution of licencing options
\item Presence and type of code URLs, also serving as prerequisite for answering $RQ_2$
\item Necessary capabilities and custom IAM rules
\item Metadata quality issues across all metrics, in particular metadata consistency
\end{enumerate}

\subsection{Code repository data collection and analysis}

To give an answer to $RQ_2$, all identified code URLs are filtered, examined and checked out for code
and configuration analysis.
The filtering distinguishes between GitHub repositories, other recognisable public repositories,
plain websites, invalid entries, and missing entries. GitHub repositories are examined for statistical information on
popularity and the dominant programming language. Moreover, deployment artefacts are extracted
to get a better picture of how cloud functions are implemented and deployed. From these,
further references to local files and archives or even remote files are followed.

A daily snapshot of all publicly accessible valid repositories is maintained updated and processed with
code and configuration analysis tools. Among them are standard tools such as SLOCcount \cite{Terceiro09structuralcomplexity}, but also
custom tools to determine specific code metrics and AWS SAM deployment entries.

The subsequent answer to $RQ_3$ is based on the evolving timeseries of all metadata and data which is stored in an
efficient way, storing and transferring only actual changes as they happen.

\section{Results}

In this section, the state of AWS SAR at the end of the one-year automated observation period is reported on,
based on snapshot day May 11, 2019. The timeline which led to this state
starting from the launch of the marketplace is also shown for selected metrics, with live update links for
evolving metrics.

\subsection{How functions are offered}

\paragraph{Data model.}
AWS SAR contains functions (more rigorously, orchestrated function-based applications) of which metadata, FaaS configuration and deployment
are described according to the AWS Serverless Application Model
(AWS SAM) specification. The original JSON configuration format captures mostly technical properties
including at package level (version numbers, dependencies) and at configuration level
(execution handler, assigned runtime memory, timeout, parameter annotations and event trigger connections).
The extended YAML deployment format introduced with SAM adds bindings to BaaS as well as nested application support
to combine multiple functions.
On top of these technical characteristics, the static publishing metadata model for AWS SAM functions in
AWS SAR contains the following JSON-formatted properties which are typically given by the function developers
upon registration in the repository:

\begin{enumerate}
\item Application name without uniqueness guarantees (e.g. \function{alexa-skills-kit-nodejs-factskill})
\item Author (publisher, e.g. Alexa Skills Kit)
\item URL (e.g. GitHub repository, home page)
\item Human-readable description (e.g. \textit{This Alexa sample skill is a template...})
\item Labels (tags, e.g. skills,fact,alexa)
\item Licencing information, README and version information -- not part of the exported data model, but linked to it
\item Required capabilities or custom identity rules -- if not present, the function will execute in unprivileged mode
\end{enumerate}

The publishing process differentiates between publicly offered functions, which most be open source
and can therefore be deployed (with evident issues) in other cloud environments, and private functions
which are only visible within one account.
After publishing a public function, this model is successively enhanced with dynamic information by the marketplace.

\begin{enumerate}
\item Fully-qualified unique resource identifier (e.g. \texttt{arn:aws:serverlessrepo:\\us-east-1:*:applications/alexa-...})
\item Number of deployments, serving as measure for popularity along with associated code repository popularity metrics
\end{enumerate}

\paragraph{Function and deployment statistics.}
As of mid-May 2019, there are 533 public functions in AWS SAR by 232 vendors, with the top vendor being 'AWS' (105 functions), followed by
an individual (29), 'AWS Secrets Manager' (15) and a long tail of various companies and individuals.
The global average supply is 2.3 functions per vendor. These numbers suggest a comparatively small community of independent Lambda developers
making use of the repository. The enforced upper bound is 100 public applications per account and region, which is exceeded only
by 'AWS'. The limit does not apply to private functions which are however not exposed through the public repository interface
and not considered in this study; hence, the number of functions registered in the marketplace overall remains unknown but can be
expected to be much higher than the number of public functions.

Of all functions, 95 (17.8\%) require special capabilities or custom rules; five functions (0.9\%) require even three capabilities
in conjunction. There is no automated way to retrieve justifications for any privileged execution, suggesting that problems
similar to privileged mobile phone applications \cite{DBLP:conf/cisc/MengBLLW18} may occur.

Correspondingly, there have been 60010 deployments of the offered functions, with the vastly dominating top deployment being
\function{alexa-skills-kit-nodejs-factskill} which alone is deployed more than all other cloud functions combined.
This top spot is followed by various similar Alexa skills such as
\function{nodejs-triviaskill} and \function{nodejs-howtoskill},
but also the obligatory \function{hello-world} function and \function{SecretsManagerRDSMySQLRotationSingleUser}.
Yet again, a long-tail distribution follows with several popular variants of database rotation functions. The top
ten deployments are shown in Table \ref{tab:deployments}.

\begin{table}[htb]
\centering
\caption{Serverless application deployment numbers in AWS SAR (Abbreviations: Depl -- Deployments; Perc -- Percentage)\label{tab:deployments}}
\begin{tabular}{p{5.2cm}lrr} \hline
\textbf{Function}				& \textbf{Vendor}	& \textbf{Depl}	& \textbf{Perc}	\\ \hline

\function{alexa-skills-kit-nodejs-factskill}		& Alexa Skills Kit	& 32434	& 54.0\%	\\ \hline
\function{alexa-skills-kit-nodejs-triviaskill}		& Alexa Skills Kit	& 2771	& 4.6\%	\\ \hline
\function{hello-world}					& AWS			& 2058	& 3.4\%	\\ \hline
\function{SecretsManagerRDSMySQL\-RotationSingleUser}	& AWS Secrets Manager	& 1971	& 3.3\%	\\ \hline
\function{alexa-skills-kit-nodejs-howtoskill}		& Alexa Skills Kit	& 1631	& 2.7\%	\\ \hline
\function{SecretsManagerRDSPostgre\-SQLRotationSingleUser}	& AWS Secrets Manager	& 1125	& 1.9\%	\\ \hline
\function{microservice-http-endpoint}			& AWS			& 763	& 1.3\%	\\ \hline
\function{alexa-skills-kit-color-expert-python}		& AWS			& 672	& 1.1\%	\\ \hline
\function{image-resizer-service}			& Cagatay Gurturk	& 602	& 1.0\%	\\ \hline
\function{SecretsManagerRDSMySQL\-RotationMultiUser}	& AWS Secrets Manager	& 558	& 0.9\%	\\ \hline
\end{tabular}
\end{table}

Overall, AWS through its multiple vendor designations
offers 26\% of all functions but due to their popularity their functions account for 84\% of all deployments,
including the top-eight functions available from the repository which alone combine to 72\%.
Comparatively, these are the outstanding averaged deployment ratios for vendors across all of their functions:

\begin{enumerate}
\item AWS or the AWS-related Alexa and Amazon brands with 50 or more deployments per function: 'Alexa Skills Kit' (7473), 'Amazon API Gateway Team' (325), 'AWS Secrets Manager' (292), 'Amazon API Gateway' (198), 'AmazonConnectSalesforceIntegration' (133), 'AWS' (73), 'Alexa for Business' (55)
\item Third-party companies or individuals with over 200 deployments per function: 'Cagatay Gurturk' (602), 'Digital Sailors' (284), 'Datadog' (288), 'evan chiu' (267) and 'Jagsp' (229)
\end{enumerate}

The global average is 113 deployments per function, and 259 deployments per vendor, which with the exception of three
AWS-designated vendors and four individuals is not exceeded by any vendor. When discarding the functions requiring capabilities,
the averages are significantly higher with 133 deployments per function and 316 deployments per vendor. In other words,
providing functions running in unprivileged mode leads to an average 18-22\% popularity boost.

\paragraph{Duplicity statistics.}
The documentation of SAR does not inform about whether function identifiers without any additional qualifiers (account name
or ARN) are supposed to be unique from a user perspective. In practice, duplicates are occasionally visible. In two such cases,
AWS-provided functions (\function{kinesis-firehose-cloudwatch-logs-processor} and \function{api-gateway-multiple-origin-cors})
are also provided
by individuals under the same name but with different licences and potentially different runtime behaviour.
In another case, a function \function{inbound-ses-spam-filter} even appeared twice offered by AWS as vendor, with identical record
including the fully-qualified identifier.

\begin{figure}[htb]
\centering
\includegraphics[width=0.8\columnwidth]{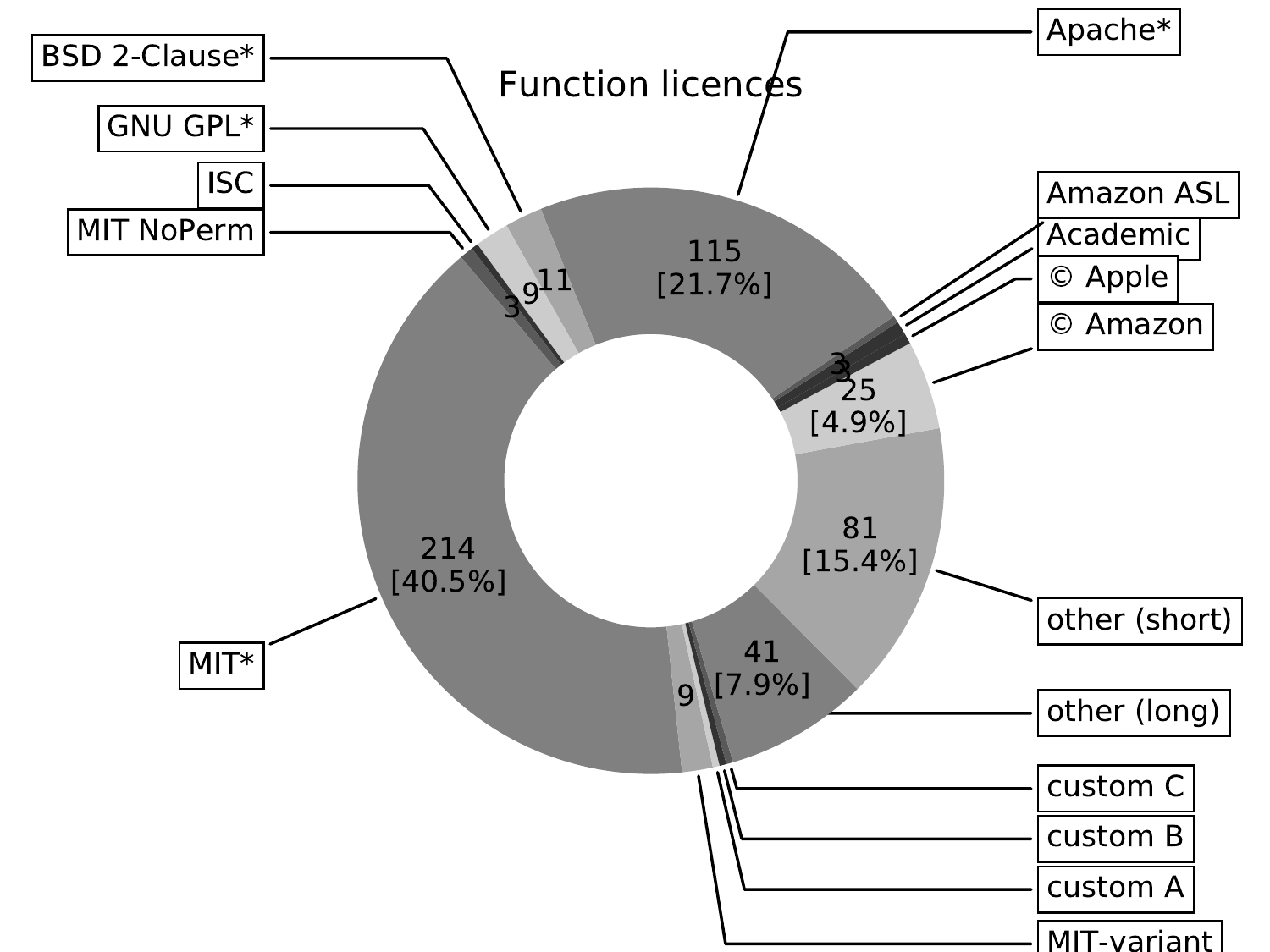}
\caption{Distribution of licences across functions\label{fig:licences}}
\end{figure}

\paragraph{Licence statistics.}
In AWS SAR, each function metadata references a mandatory licence text and a longer README file. The analysis shows that
uniformity is not enforced.
Out of 531 licences, the dominating ones (including variants) are: MIT Licence (40.5\%), Apache Licence (21.7\%) and
the proprietary Amazon Licence (4.9\%), ahead of the otherwise traditional GNU GPL (1.9\%) which, surprisingly,
contains one instance of the 30-year old GPLv1 (\function{Voice-Lexicon-API} function). A total of 15 licences
occurs more than once. In contrast, 7.9\% of licence texts are unique legal texts, and another 15.4\% are unique short texts not
corresponding to actual licencing information but rather containing placeholder text.
Fig. \ref{fig:licences} shows the distribution of licences in a chart.

\paragraph{Documentation statistics.}
From a software developer perspective, finding the right function quickly is highly important. An efficient function
search requires appropriate tags, high-quality brief descriptions and extensive documentation.

Among all functions, 417 (78.2\%) are tagged while 116 (21.8\%) are not. In total, there are 1914 tags which corresponds
to an average of 3.6 tags per function. Considering that there are 750 unique tags in total, each tag is on average re-used
less than 2.6 times which indicates that most tags are not currently useful to search for candidate functions.
The most-used tags with at least 30 occurrences are predominantly AWS-specific terms such as 'lambda' (92 times), 's3' (59 times), 'nodejs', 'AWS', 'api', 'Lambda' and 'dynamodb'. On average, a tag is 7.3 characters long, while some are more descriptive and up to 29 characters long (\function{salesforce-api-access-manager}).

\begin{figure}[H]
\centering
\href{https://raw.githubusercontent.com/serviceprototypinglab/aws-sar-dataset/master/plots/insights-longtail-tags.summer.png}{\includegraphics[width=0.8\columnwidth]{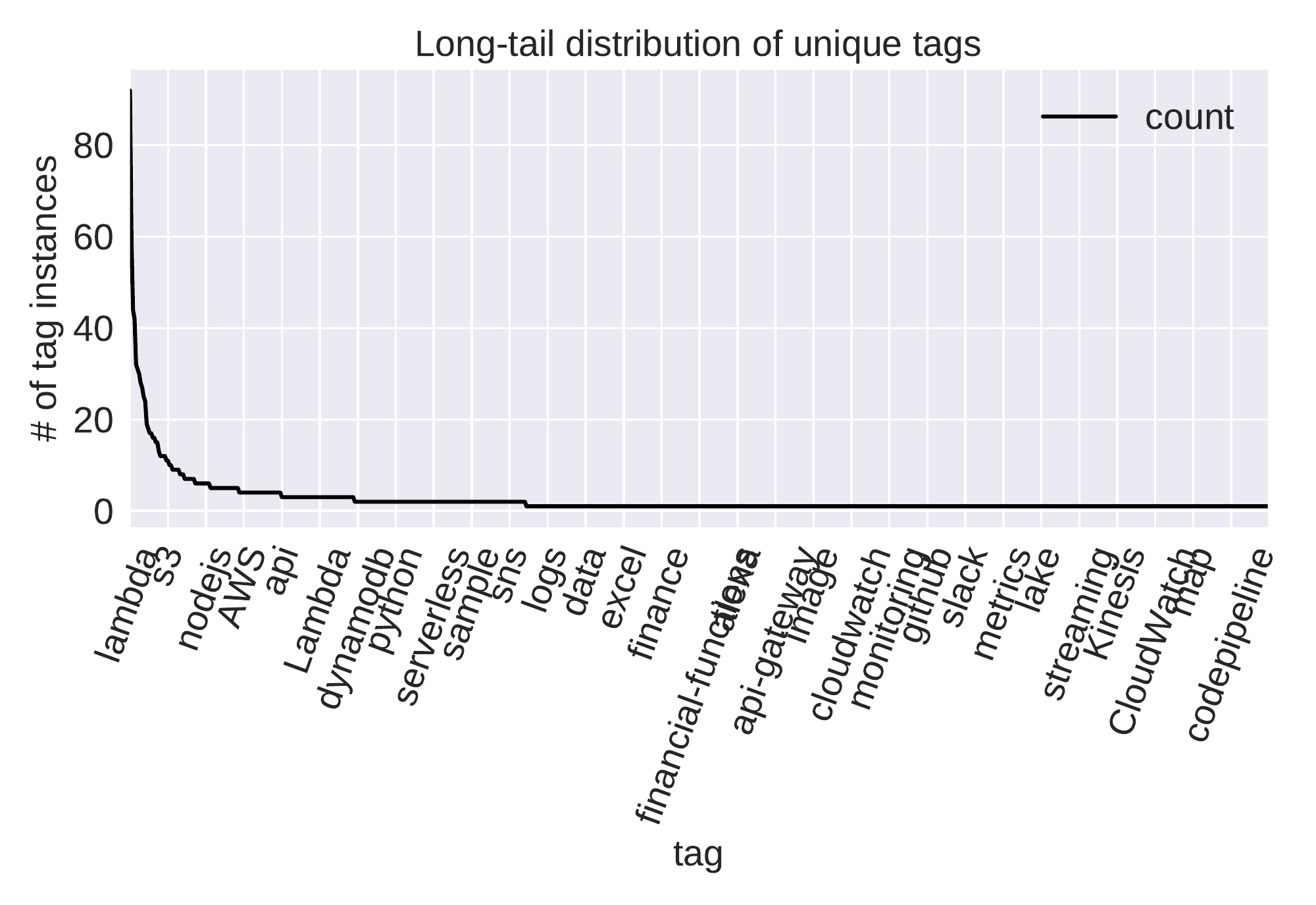}}
\caption{Distribution of tag instances per unique tag (\liveupdate)\label{fig:longtails}}
\end{figure}

The long-tail distribution of selected tags across the frequency spectrum on cloud functions is shown in Fig. \ref{fig:longtails}.
Additionally, Table \ref{tab:tags} classifies the 15 most-used tags according to whether they describe a generic technical
subject or technology or a vendor-specific service, referring to dependencies within the AWS ecosystem.
Evidently, the most-used tags are not descriptive enough for functional
discovery but rather used as filter for technology alignment.

\begin{table}[htb]
\centering
\caption{Most-used tags across functions\label{tab:tags}}
\begin{tabular}{lrcc} \hline
\textbf{Tag}	& \textbf{Frequency}	& \textbf{Generic Term}	& \textbf{Vendor Term}		\\ \hline

lambda		& 90			&			& X				\\ \hline
s3		& 59			&			& X				\\ \hline
nodejs		& 44			& X			&				\\ \hline
AWS		& 42			&			& X				\\ \hline
api		& 32			& X			&				\\ \hline
Lambda		& 31			&			& X				\\ \hline
dynamodb	& 30			&			& X				\\ \hline
python		& 28			& X			&				\\ \hline
serverless	& 27			& X			&				\\ \hline
sample		& 25			& X			&				\\ \hline
sns		& 24			&			& X				\\ \hline
logs		& 19			& X			&				\\ \hline
data		& 18			& X			&				\\ \hline
finance		& 17			& X			&				\\ \hline
excel		& 17			& X			&				\\ \hline

\end{tabular}
\end{table}

The short description text for Lambda functions is limited to 256 characters. Consequently, description texts vary
almost linearly from few characters such as 'This is a test' and around 20 similarly unexpressive descriptions to the permissible maximum
such as 'This is a serverless component which sends an email to specified email addresses...'. Interestingly,
one of the descriptions is in Japanese while all others are in English, and one is a link to a website.
The similarity ratio among all descriptions, applicable to all text strings with 90\% overlap or more, is 14.1\%.
This number suggests that around one out of seven functions is a feature or implementation variant of another one.

With a range comparable to that of licences and short descriptions, the mandatory README files range from few-character placeholders to
extensive texts with multi-section markdown structure of around 10-12 kB.
A large outlier is \function{applicationName-227c1372-76ee-4797-a593-83d19dc4f264}
provided by 'author', arguably a test entry, whose README contains an entire web page of around 250 kB.
This observation suggests that while AWS confirms manual code quality and licence conformance checks,
there are no such checks on metadata or documentation, and the checks on licences do not prevent the emergence
of a licence jungle.

\subsection{How functions are implemented}

\paragraph{Code repository information.}
Among all functions, 455 (85.4\%) specify a URL with further information. Sometimes, this carries the semantics of a
semi-structured home page without direct reference to the implementation, but more often, applying to 392 functions (73.5\%),
it refers to a GitHub repository which can be analysed automatically. The number of unique repositories is 255, which implies
that several functions share one repository, in fact up to 16 of them. The shared repository links are either
identical, so that finding the corresponding implementation and configuration becomes heuristic, or different
by path and/or branch so that a 1:1 mapping of function to implementation location remains possible.
Among the non-GitHub links, the majority point to
the vendor sites 'aws.amazon.com' (13) and 'www.streamdata.io' (11), while most others have only singular or double
occurrence. The dominance of GitHub, beside the popularity of
the platform itself, can be explained with the streamlined code publishing process in \service{AWS CodePipeline} which updates
cloud functions in AWS SAR with every Git commit.

Moreover, some URLs are invalid either syntactically
or by referring to non-existing or no longer available content, to password-protected private repositories or to collections of repositories
rather than a single repository. Hence, of the 392 GitHub repositories, only 365 (93.1\%)
could be eventually assessed; among the unique repositories, discarding differences in just the paths, only 238 (93.3\%),
consequently containing 374 cloud functions in 325 different paths or branches.
Even less URLs point properly to the directory containing only the relevant function code,
requiring further manual work before an eventual code analysis and leading to unnecessary transmission
of repository overhead.

\paragraph{Code repository characteristics.}

The aggregate size of the repositories is 2.4 GB. For practical reasons of copying referenced paths and files and the ability to 
employ external statistics tools, a checked-out copy weighs in at 6.8 GB.
Extracting all directories referenced from function metadata, and removing any superfluous versioning information,
produces a self-contained implementation folder with a net size of 4.4 GB.

The largest repository by far is \repo{lambda-packs} which contains 21 pre-compiled Lambda functions, such as
\function{Tesseract}, \function{Chromium} and \function{Pandas}, including
dependency libraries. Eight of these functions reference the entire pack collection, contributing to a high
and redundant capacity requirement which is avoided by the compact duplicate-considerate representation.
The repository with most functions is \repo{awslabs/serverless-application-model}, containing the popular \function{hello-world}
function along with many others, for a total count of 106 including variants and tests, of which 83 are published
on SAR under the AWS and AWS Greengrass vendor designations.

The code repositories show a skewed GitHub popularity distribution with significant factors (around 20 to 80) between the mean
and median values. The overall top repository (\repo{hello-world} and its variants) has 5146 stars,
326 watchers and 1183 forks.

\paragraph{Programming languages.}
Programming languages are assessed on three levels: code repository metadata, code files, and Lambda configurations.

On the code repository metadata level, the most popular languages are indicated to be JavaScript with 108 functions (29.6\%) and Python
with 206 functions (56.4\%), followed by 10 functions each with Go and Java code (2.7\% each).

Relative to the assessable code bases spread across 322 unique folders,
159 functions directories (50.0\%) contain JavaScript code and 140 (44.0\%) contain Python
code, irrespective of the code size or complexity, revealing a slightly different picture mostly caused by shared repositories.
Moreover, repositories contain a substantial amount of maintenance languages such as Shell code (153 or 48.1\%) and Perl (117 or 36.8\%).
Minor languages with occurrence in less than 10 directories according to both statistics are Java and Go.
Despite the theoretic possibility of multiple languages being used per repository, this is hardly the case according to the observation.
Hence, polyglot implementations are not common
and despite increasing language support by all FaaS providers, developer interest beyond JavaScript and Python is limited.
This confirms previous empirical findings \cite{faasproductionjournal}.

Relative to the FaaS runtimes specified in the Lambda configuration files, often more than one per directory due to composite
functions or variants, 305 functions out of 587 (52.0\%) use different versions of JavaScript,
214 (36.5\%) different versions of Python, and very few indicate either another runtime (5.6\%) or none (6.0\%).

Concerning the functionality, the complexity ranges from multi-functional function suites such as Serverless
Galeria for image processing to very simple and even profane functions such as a 'daily doggo' picture
submission service and a nude filter.

\begin{figure}[htb]
\centering
\href{https://raw.githubusercontent.com/serviceprototypinglab/aws-sar-dataset/master/plots/sloc.png}{\includegraphics[width=0.8\columnwidth]{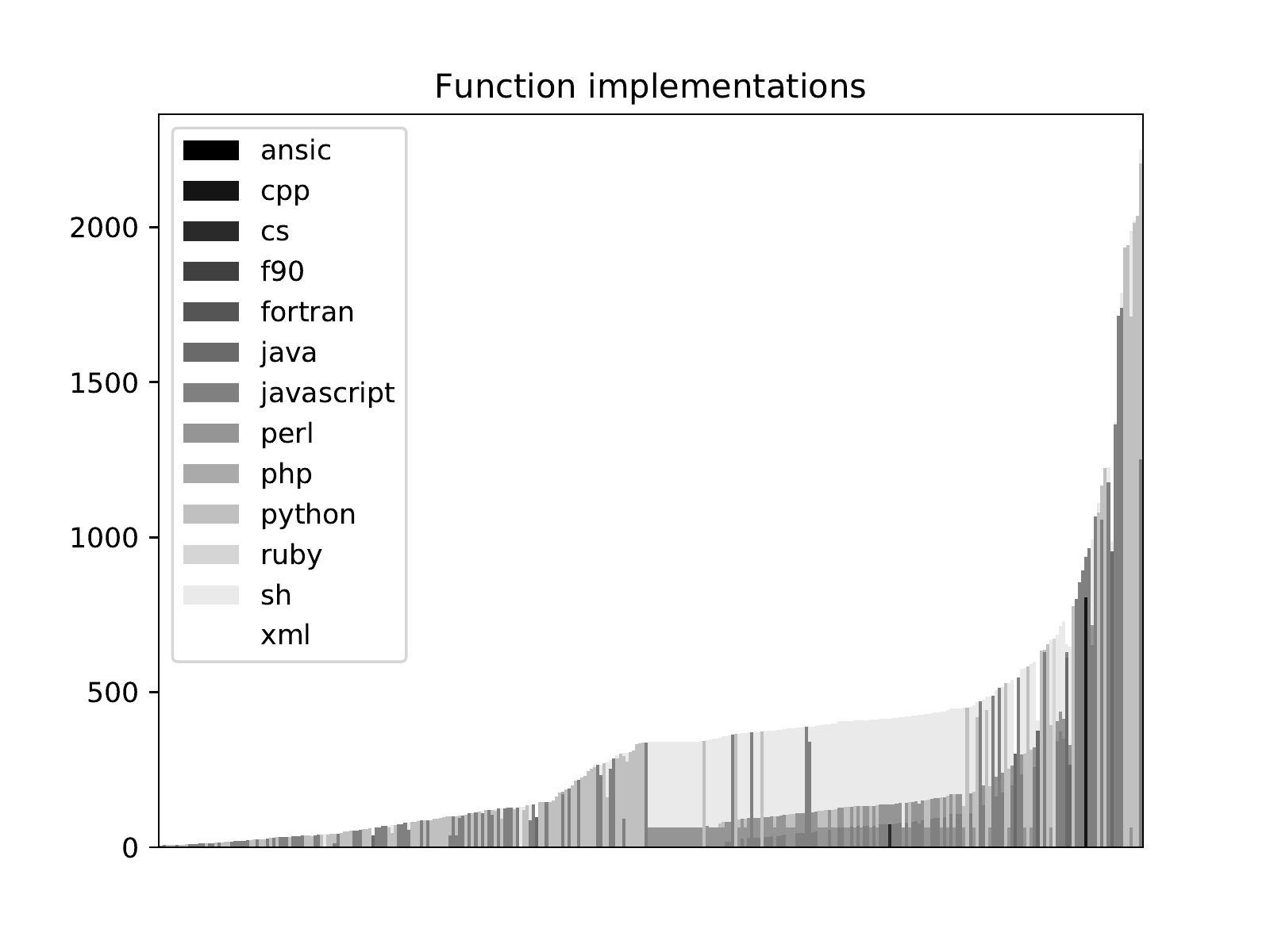}}
\caption{Programming languages and code sizes (\liveupdate)\label{fig:proglang}}
\end{figure}

Fig. \ref{fig:proglang} summarises the programming languages according to the code files analysis. Repositories with more
than 3000 lines of code are excluded for reasons of visualisation, leaving 306 out of 322 unique code folders (95.0\%)
while excluding among others the largest repository with over 8 million lines of predominantly Python code, most of which
is not representative of the actual cloud function.
The diagram shows a large similarity of the middle section in which more than half of the code consists of maintenance
shell scripts while the distinct bars show JavaScript (darker grey) and Python (brighter grey).

\paragraph{Programming models.}
This study provides only a high-level sample peek into the code structures and programming models of the cloud functions themselves.
The reason for this limitation is the large diversity in terms of structures, models and patterns, and the need to
develop custom metering tools for each programming language, in conjunction with the nondeterministic mapping of code
repository contents to functions. Nevertheless, the sampling already confirms some previously empirically confirmed patterns
such as dispatcher functions \cite{faasproductionjournal}. Table \ref{tab:code} shows a sample of ten different functions
from the vendor 'AWS'. Its repository folder identifiers are taken from the reference dataset.

\begin{table}[htb]
\centering
\caption{Structural Lambda code characteristics\label{tab:code}}
\begin{tabular}{lllll} \hline
\textbf{Id\#}	& \textbf{Language}	& \textbf{Dependencies}	& \textbf{Files}	& \textbf{Code}	\\ \hline

3-102	& python	& requests	& 1 & "dispatcher", 21 functions	\\ \hline
3-103	& nodejs	& algorithmia	& 1 & "simple", 2 nested callbacks	\\ \hline
3-107	& python	& requests	& 1 & "simple-iterator"			\\ \hline
3-108	& mixed		& (algorithmia)	& 2 & "simple", global variables	\\ \hline
3-113	& nodejs	& algorithmia	& 1 & "simple", 1 callback		\\ \hline
3-114	& python	& --		& 1 & "simple-iterator", 5 methods	\\ \hline
3-116	& python	& requests	& 1 & "dispatcher", 21 functions	\\ \hline
3-118	& python	& requests	& 1 & "simple", 2 functions		\\ \hline
3-121	& python	& requests	& 1 & "dispatcher", 10 functions	\\ \hline
3-127	& nodejs	& --		& 1 & "simple", 3 functions		\\ \hline
\end{tabular}
\end{table}

Only one out of ten cloud function implementations uses object-oriented programming.
Noteworthy is the popularity of the
\filename{requests} module which suggests that it could be included in Lambda's Python runtime by default.
Repository folder 3-108 (function \function{cloudwatch-alarm-to-slack-python}) contains both a Python and a JavaScript
implementation of the same function, presumably due to a programming mistake by copying \function{cloudwatch-alarm-to-slack}
without removing the original code file. All other folders contain a single source file.
The code analysis shows further sources for duplicity, such as almost-identical functions with slightly different
syntax to target distinct versions of a programming language, for instance, Python 2.7 and Python 3.6.

\paragraph{Function application orchestration.}
The exploration of SAM files allow for a better understanding of function implementation retrieval and
the composite nature of Lambda-based applications. In total, there are 509 SAM files or on average 1.6
per code folder. Finding them is a heuristic process due to different names, although many are called
\filename{template.yaml}, and due to the co-existence with other YAML files in many software projects.

285 code repositories (88.5\%) contain SAM files which configure the Lambda execution and reveal bindings to BaaS through
resource definitions.
Among all SAM files, 587 resources of type \syntax{AWS::Serverless::Function} are defined, averaging
more than one per file and covering 272 (95.4\%) of all code folders. Beyond this dominant resource entry,
several BaaS entries stand out albeit all at lesser scale. The top resources are \syntax{AWS::S3::Bucket}
(103, 20.2\%), \syntax{AWS::IAM::Role} (61, 12.0\%), \syntax{AWS::Serverless::Api} (45, 8.8\%)
and \syntax{AWS\-::DynamoDB::Table} (44, 8.6\%). This suggests that the typical Lambda-based application stores
blob data on \service{Amazon S3}, relational data in \service{DynamoDB}, and is invoked externally through the \service{API Gateway}.

A clustering analysis discarding the multiplicity of resource types reveals common structures of serverless applications
as precursor to a deeper cloud function pattern analysis which has recently become a research topic \cite{SLD_58}.
There are 93 distinct resource composition types across all SAM files of which nine can be considered significant due to occurring
more than five times each. Table \ref{tab:samclusters} shows these patterns, among which the single function is occurring
in every second serverless application. This does not necessarily mean that no BaaS is involved in such applications,
but it does depart with the view that in practice, serverless applications are mostly expressed as explicit composition
of a FaaS resource with one or more BaaS resources.
Among the most-popular functions, \function{Alexa Facts Skill} has its data hard-coded in the implementation and
\function{Alexa Trivia} and \function{HowTo Skill}s
load data from local files. Despite Lambda resource limits, such almost monolithic designs appear to be popular with developers.

\begin{table}[htb]
\centering
\caption{Serverless application composition types in AWS SAR\label{tab:samclusters}}
\begin{tabular}{lrr} \hline
\textbf{Type}				& \textbf{Occurrences}	& \textbf{Percentage}	\\ \hline

Function				& 269	& 53.0\%\\ \hline
Function + S3 (storage)			& 40	& 7.9\%	\\ \hline
Function + API-Gateway			& 25	& 4.9\%	\\ \hline
Function + SNS (notifications)		& 19	& 3.7\%	\\ \hline
Function + SimpleTable (database)	& 13	& 2.6\%	\\ \hline
Function + DynamoDB (database)		& 11	& 2.2\%	\\ \hline
Function + Permission			& 8	& 1.6\%	\\ \hline
Function + Kinesis (streaming)		& 6	& 1.2\%	\\ \hline
Function + IAM (authorisation)		& 6	& 1.2\% \\ \hline
\end{tabular}
\end{table}

On the other end of the spectrum, individual compositions with only one occurrence but multiple resources exist. The most complex composition, named \function{api-gateway-dev-portal}, involves 17 resource types including the AWS cloud services CloudFront, Cognito, DymamoDB, Route53, S3, SNS, API Gateway and IAM, as well as custom resource policies and other capabilities.

Among all SAM files, 69 reference no code, 63 reference only remote (S3-hosted) code, 4 reference local code encapsulated
in ZIP files, and 373 reference repository-local code paths. The definitions without code are in most cases either
composite applications or pre-configured blank templates in which developers still insert code but are otherwise
already readily connected to BaaS.
By excluding code repositories which reference no or only remote data,
the cumulative code size shrinks from 4.4 GB to a mere 585 MB, making a further code analysis easier.

\begin{figure}[htb]
\centering
\includegraphics[width=1.0\columnwidth]{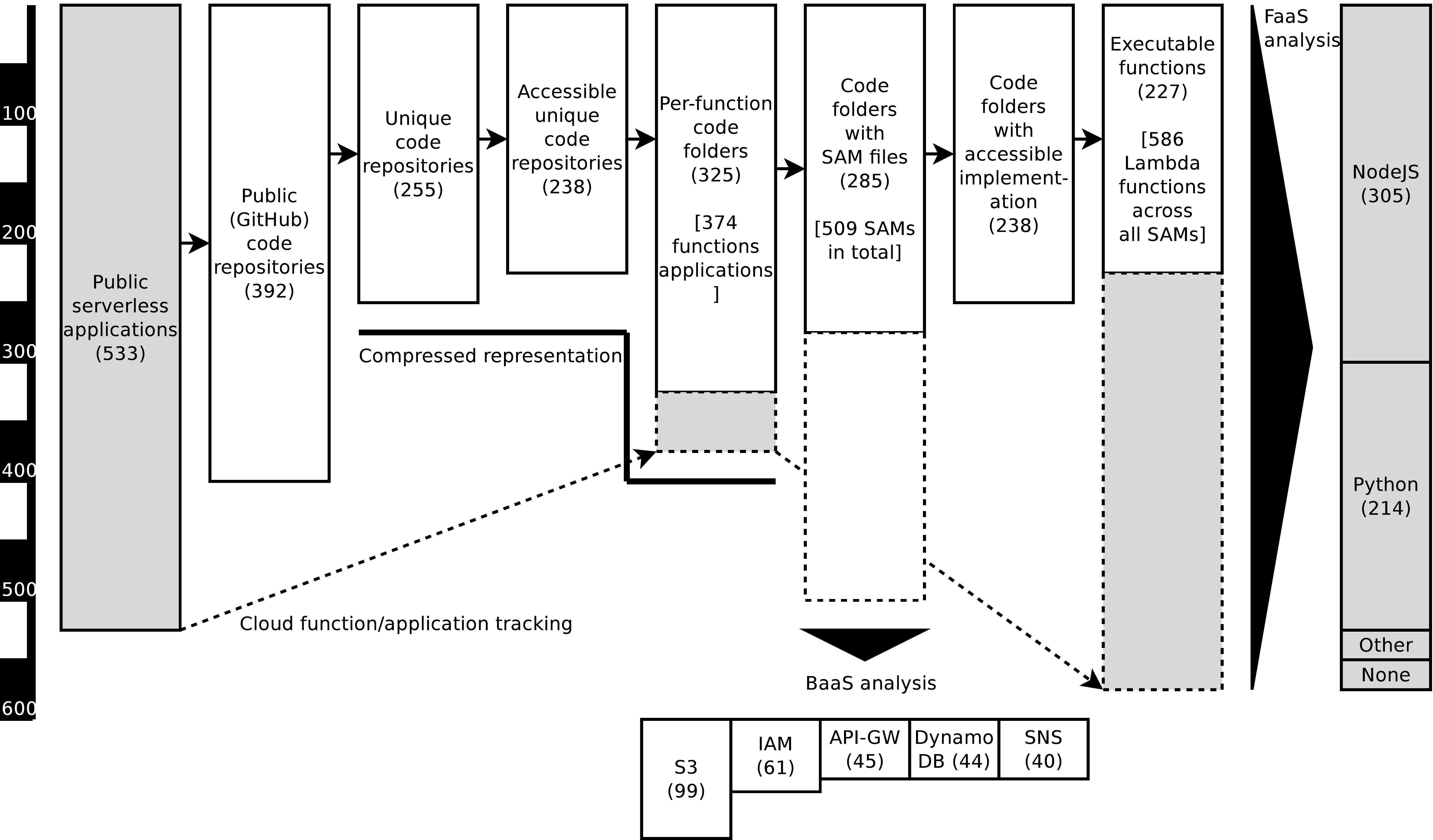}
\caption{Code drill-down process from SAR to FaaS and BaaS metrics\label{fig:codedrilldown}}
\end{figure}

\paragraph{Entities, relations and numbers.}
The entire drill-down process from the AWS SAR metadata to code repositories, function-specific directories,
SAM files and external implementation references is shown in Fig. \ref{fig:codedrilldown}.
Eventually, the rather complex constructs lead to key metrics on the use of FaaS and BaaS such as
occurrence statistics and patterns.

\subsection{Which change patterns exist}

\begin{figure}[htb]
\centering
\href{https://raw.githubusercontent.com/serviceprototypinglab/aws-sar-dataset/master/plots/autostats-plot.summer.png}{\includegraphics[width=0.8\columnwidth]{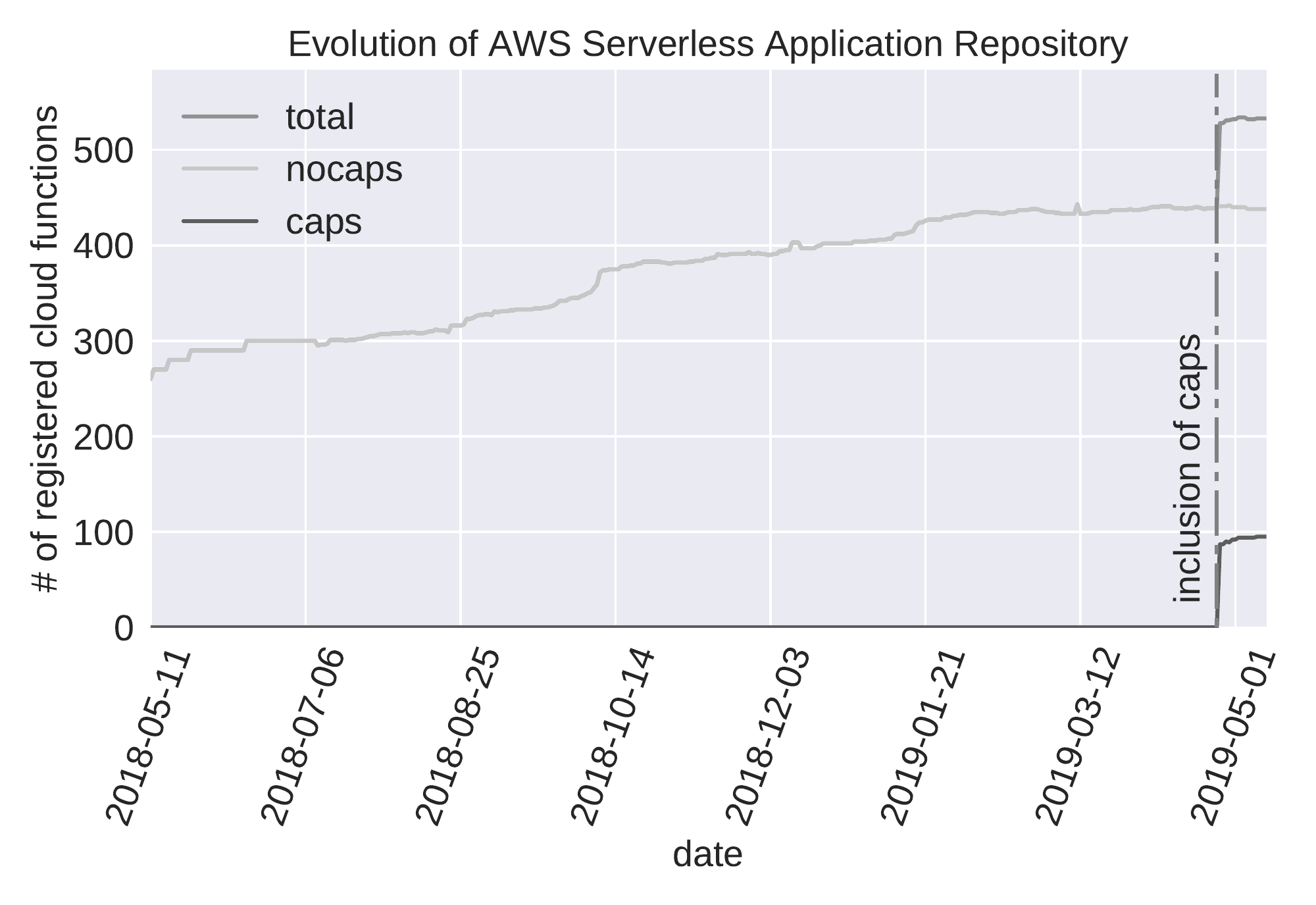}}
\caption{Number of cloud functions at AWS SAR over time (\liveupdate)\label{fig:autostats}}
\end{figure}

\paragraph{Evolution of functions and deployments.}
At launch time on February 28, 2018, AWS SAR contained already 180 functions, based on approximated manual observation.
This number grew to approximated 260 on mid-May 2018. The further development captured through automated observation is shown in Fig. \ref{fig:autostats},
first still approximated until mid-July, 2018, and afterwards, due to more precise measurements, as actual
numbers, with privileged functions omitted from the measurements until late April 2019.
After one year of automated observation, 438 non-privileged and 533 overall public functions (with "caps"
meaning required deployment capabilities) have become available.
Overall, a stable average growth of around 14-16 functions per month is visible, corresponding
to a declining month-over-month growth more than 6\% to less than 1\%, averaging at slightly below 4\%.

In comparison, other polyglot microservice artefacts show a similar growth rate, with slightly above 4\% for Helm charts over
the same observation period.
Long-established repositories for programming language-specific artefacts however show slower monthly growth rates,
such as PyPI for Python libraries (2.4\%), Maven for Java libraries (1.1\%) and Ruby Gems (0.5\%).

\begin{figure}[htb]
\centering
\href{https://raw.githubusercontent.com/serviceprototypinglab/aws-sar-dataset/master/plots/plusminus.summer.png}{\includegraphics[width=0.8\columnwidth]{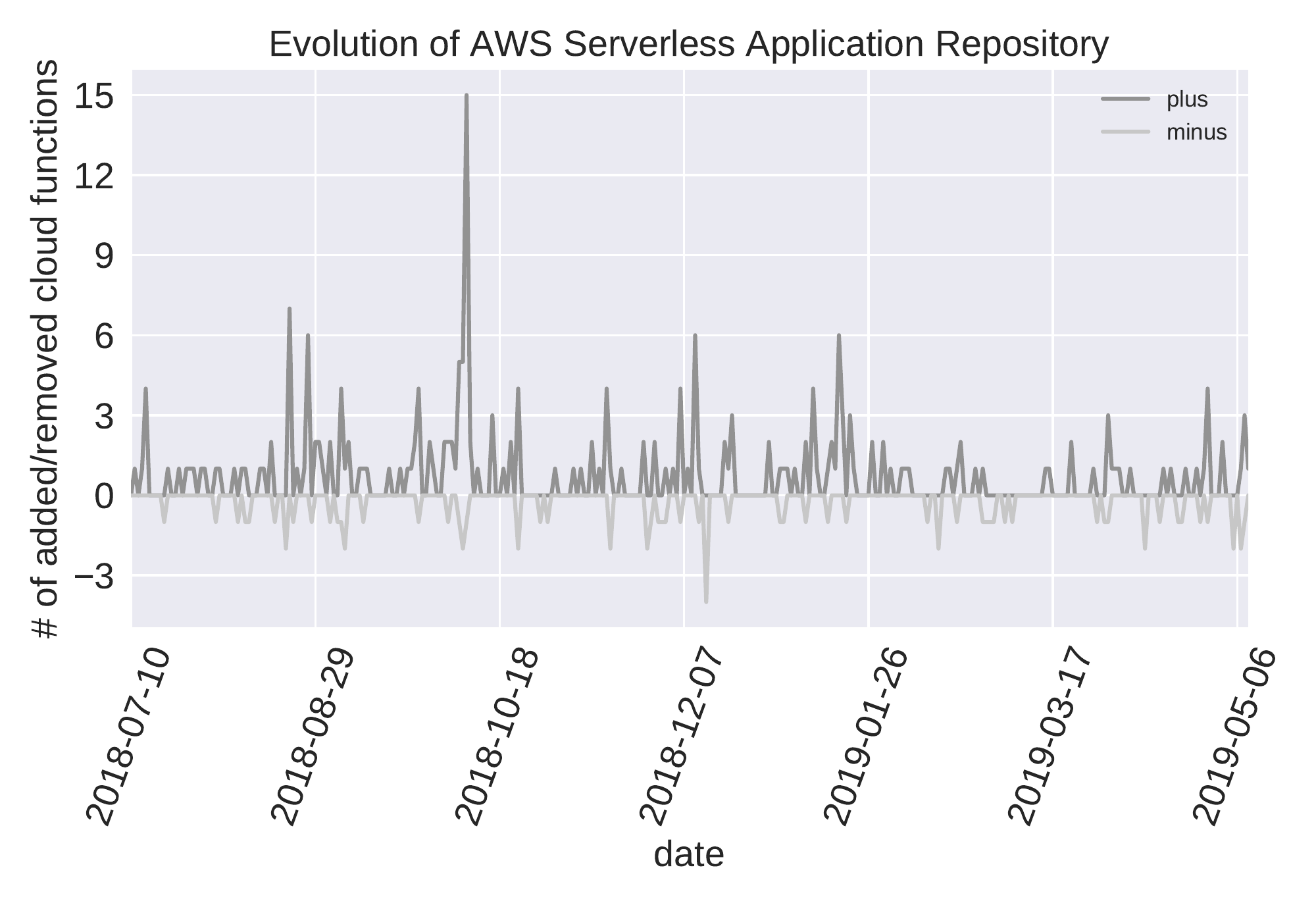}}
\caption{Volatility of cloud functions at AWS SAR over time (\liveupdate)\label{fig:plusminus}}
\end{figure}

An important complementary metric to the growth is the volatility, because the growth curve alone does not express
neutralising additions and removals of cloud functions.
The volatility with daily additions and removals is shown in Fig. \ref{fig:plusminus}. Over a period of 304
days, around 221 functions were added and 71 removed, a damping in the potential growth rate of around 24\%.
Each day, an average of 0.7 additions and 0.2 removals occur, with few changes over time to these numbers
and only few spikes mostly related to mistaken duplicate entries.
The reasons for function removals are not clear and might include renaming as well as intermittent depublication.

While the growth and decline in the number of functions represents the supply side, function deployments signal the corresponding demand side
and are therefore analysed in the same context.

\begin{figure}[H]
\centering
\href{https://raw.githubusercontent.com/serviceprototypinglab/aws-sar-dataset/master/plots/insights-percent-plot.summer.png}{\includegraphics[width=0.8\columnwidth]{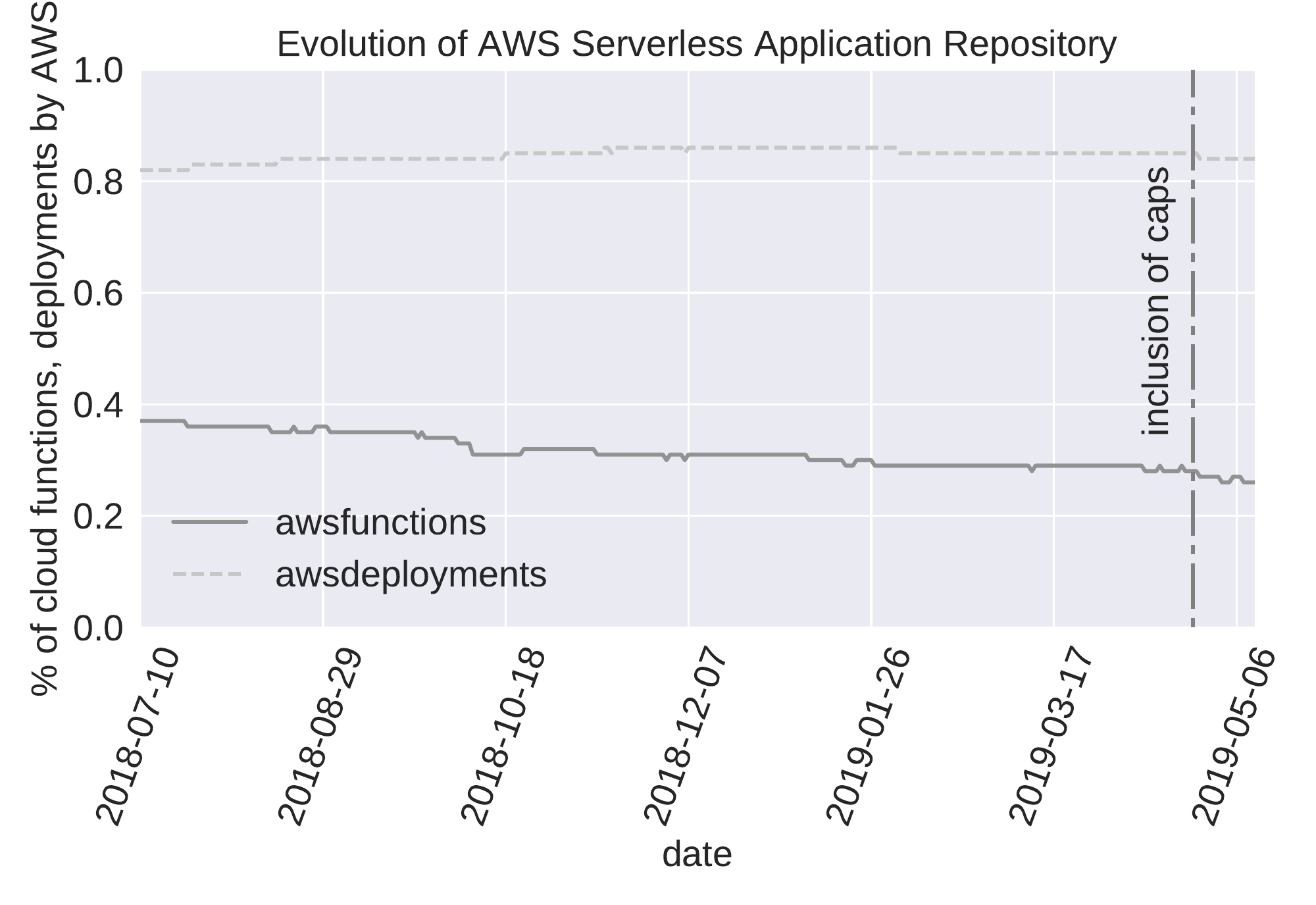}}
\caption{Ratio of AWS-provided and third-party-provided functions and their deployments over time (\liveupdate)\label{fig:percentplot}}
\end{figure}

Notably, despite a similar growth in AWS SAR vendor diversity, and in consequence a decreasing
share of AWS-provided functions among all vendors, the share of AWS-provided function deployments
has been initially increasing with almost opposite tenacity before remaining stable, as evidenced in Fig. \ref{fig:percentplot}.
Much of this growth can be attributed to the popularity of Alexa functions.
Again, the figure includes privileged functions from end-April 2019, visualised by a small sudden reduction of the
shown ratios.

\begin{figure}[H]
\centering
\href{https://raw.githubusercontent.com/serviceprototypinglab/aws-sar-dataset/master/plots/growth.gray.png}{\includegraphics[width=0.75\columnwidth]{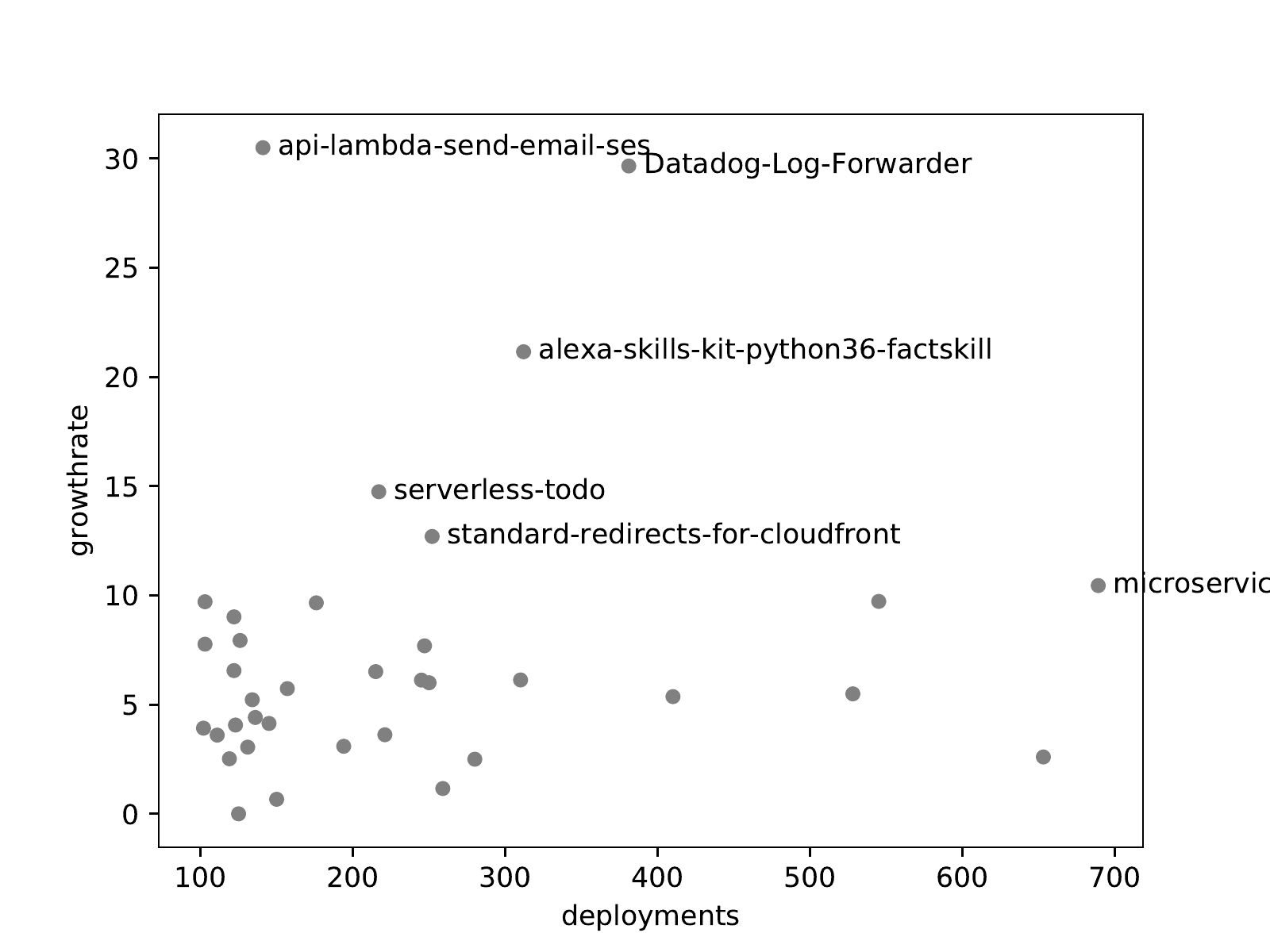}}
\caption{Single-month growth rates (in \%) of a relevant subset of functions (\liveupdate)\label{fig:growth}}
\end{figure}

A peek into trends in serverless computing is permitted by extracting trending functions, defined as fastest-growing
functions over the last month of the observation period. Evidently, the growth rate is infinite for newly appeared
functions, and still very high for functions just published before the trend window or with just few initial
deployments. Therefore, the analysis focuses on functions with already at least 100 deployments at the beginning of the
trend window. Moreover, the growth rate is typically stabilising (although not necessarily low) for widely deployed functions,
leading to an upper bound
of 1000 deployments. Both bounds are arbitrarily chosen but allow for visualising the trends. Fig. \ref{fig:growth}
shows a cluster analysis of deployment numbers to growth rates.

According to the analysis, five Lambda functions are very popular
with growth rates of 10\% or more per month, higher than the overall average of 7.0\%.
But even among the selected 35 functions, some (e.g. \function{signalfx-lambda-example-nodejs}) show zero or near-zero growth.
Among the overall top-five functions in terms of at least 1500 deployments, the growth rates are between 3.8\% and,
for the ubiquitous \function{hello-world} function, 13.4\%.

\begin{figure}[htb]
\centering
\href{https://raw.githubusercontent.com/serviceprototypinglab/aws-sar-dataset/master/plots/growthvendors.gray.png}{\includegraphics[width=0.75\columnwidth]{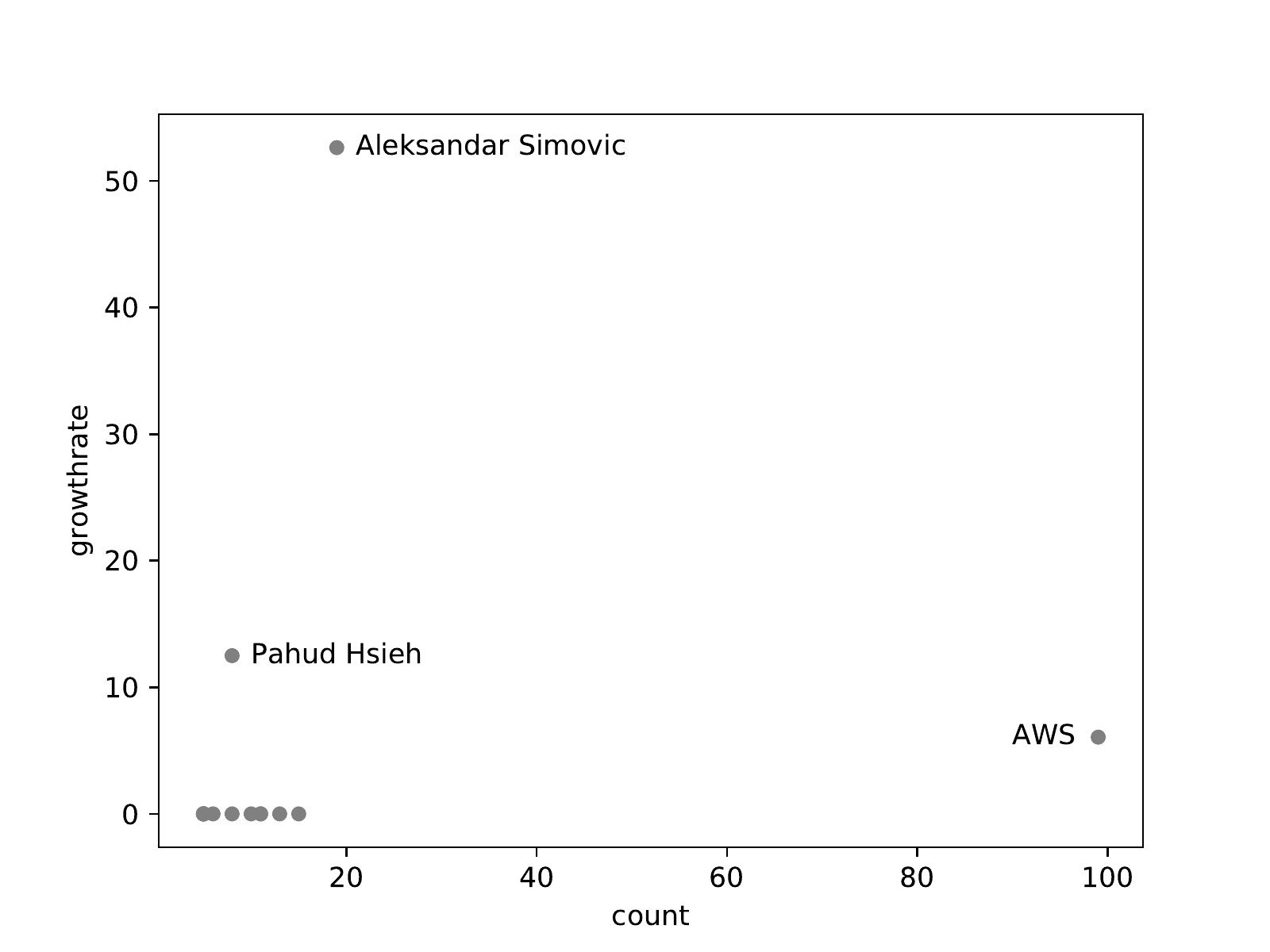}}
\caption{Single-month growth rates (in \%) of a relevant subset of vendors (\liveupdate)\label{fig:growthvendors}}
\end{figure}

The growth rate can be similarly determined for the supply side in terms of the number of functions offered by vendors.
Fig. \ref{fig:growthvendors} shows an excerpt of growth metrics for all vendors with at least five functions
in their portfolio at the beginning of the trend window. While most vendors show no growth over one month, two individuals
as well as AWS stand out for significant additions, demonstrating isolated supply-side popularity with selected
serverless application developers.

\begin{figure}[H]
\centering
\includegraphics[width=0.8\columnwidth]{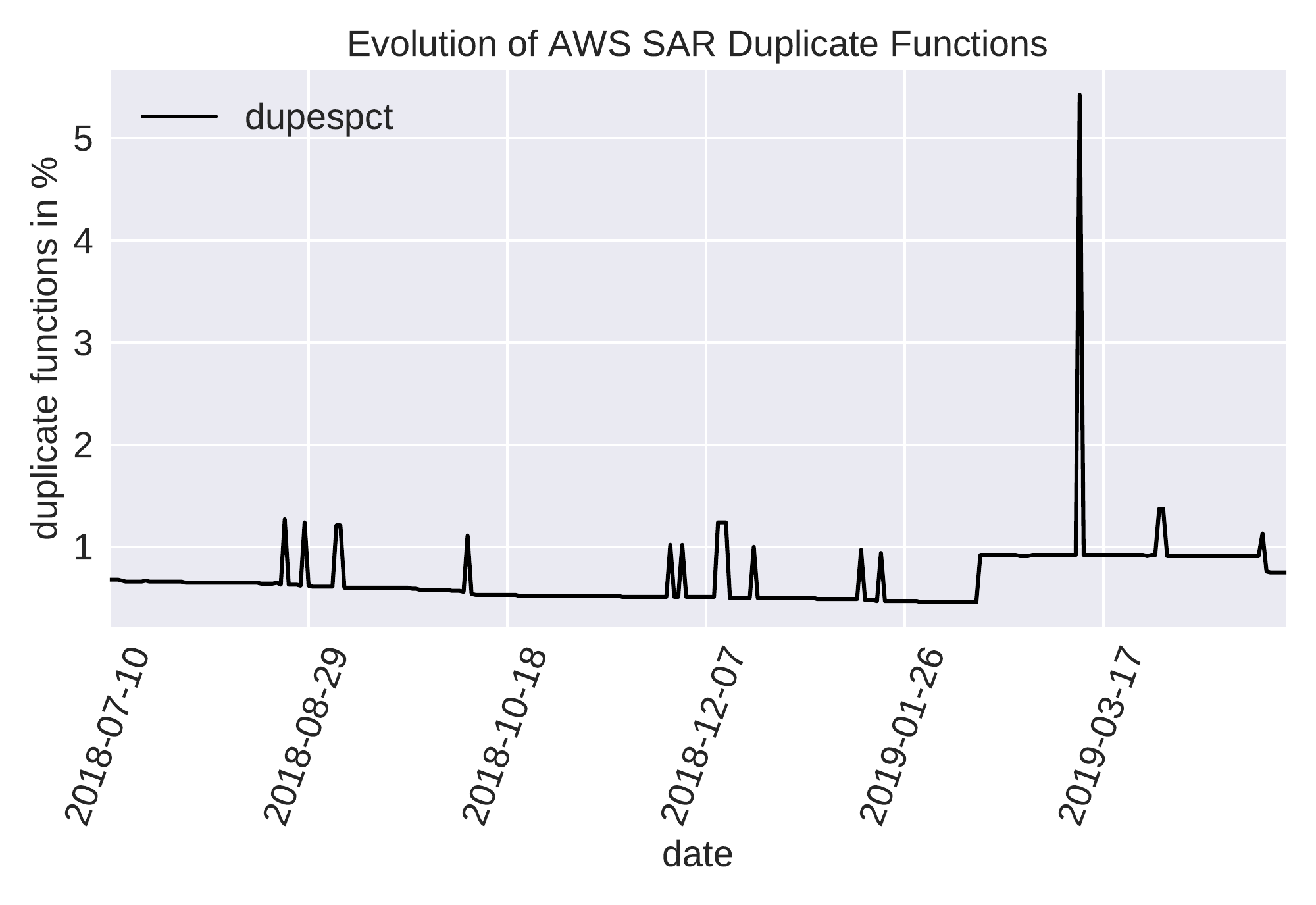}
\caption{Percentage of functions affected by duplicate names over time\label{fig:duplicates}}
\end{figure}

\paragraph{Evolution of metadata quality.}
Fig. \ref{fig:duplicates} shows the timeline of functions with duplicate names.

Evidently, while duplicates are not
prohibited by the SAR data model, they appear to result from copy and paste of function metadata by developers and are undesirable. Accordingly, upon occurrence
they are quickly corrected in most cases.
In March 2019, several functions were added twice by the vendor HERE Technologies, causing a larger spike
which statistically affected every 20th function in the repository.
Moreover, a baseline of at least two continuous duplicates remains, although it is negligible in practice as it affects
only 1\% of all offered functions.

\begin{figure}[htb]
\centering
\href{https://raw.githubusercontent.com/serviceprototypinglab/aws-sar-dataset/master/plots/codechecker.summer.png}{\includegraphics[width=0.8\columnwidth]{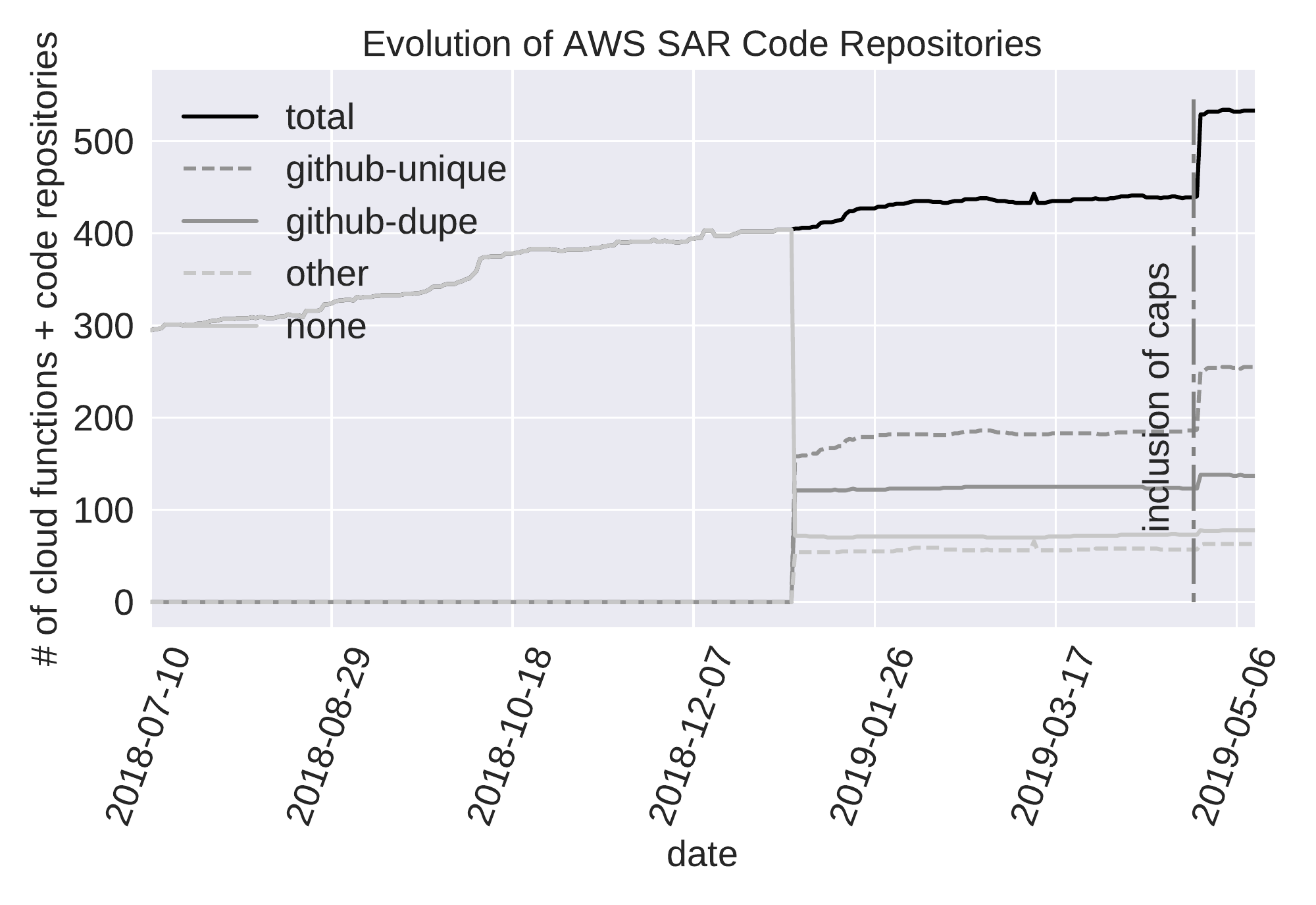}}
\caption{Code repositories linked to cloud functions (\liveupdate)\label{fig:codechecker}}
\end{figure}

\paragraph{Evolution of code repositories.}
The evolution of code repositories associated to cloud functions starting in January 2019 is shown in Fig. \ref{fig:codechecker}.
Interestingly, the growth period at the beginning of the year could be contributed almost exclusively to new
repositories rather than additional functions from existing repositories.

When comparing the number of repository forks with the deployments of cloud functions, one remarkable development
is that forks exceed deployments. This means that repositories of many cloud functions are often cloned,
perhaps in a hoarding manner. Only for the most popular Alexa functions, the number of forks can be explained
by the popularity indicated by deployments, such as developers initially deploying the original code
and then performing modifications or adding debugging statements in troubleshooting scenarios.
Private deployments may also be a reason.
A particular counterexample is \function{s3-presigned-url} which is forked a lot (1169 times) but hardly deployed (34).
In this case, the discrepancy is due to the repository sharing. Hence, the n:1 mapping of cloud functions
to repositories presents difficulties in deriving generalised statements on popularity.

\section{Discussion}

The study opens broader discussion potential on a number of topics. Three of them --
re-use, vendors, and description quality -- shall be discussed briefly here.

\paragraph{Function re-use.}
Despite the current high industry interest around cloud functions and especially around \service{AWS Lambda} as one of the
most-used services, the re-use potential of cloud functions needs to be explored in more detail.
The almost monotonic but slow growth of Lambda functions in AWS SAR suggests that most software applications are not
quickly rewritten to make use of functions, or that the resulting functions are not shared on open repositories, hubs and marketplaces.

\paragraph{Function vendors.}
The fact that AWS operates both the SAR marketplace and many of the functions resembles the schema
known from the Amazon marketplace, operated by its parent company, in which market share in products
is increasingly sought by offerings by Amazon under different names. In AWS SAR, such offerings
are openly named after AWS or well-known Amazon products such as Alexa;
still, the longer-term implication of having the double role advantage, and its effect on the large share of deployments,
will require more economic-analytical work.

\paragraph{Function description quality.}
Even though the number of functions in SAR is still manageable, several data inconsistencies are evident.
Functions with custom capabilities appear in the feed reserved for functions not requiring those (2 out of 10 of \syntax{CAPABILITY\_AUTO\_EXPAND}), vendors designations are not always clear (Amazon WorkMail vs. AWS WorkMail), function names are not unique across vendors, documentation quality varies significantly, code repository links are outdated or do not refer to code repositories at all, and licence information consists of a mix of actual licence text and placeholder text. In order to increase the automation potential for enacting and using cloud functions dynamically, small inconsistencies are avoidable obstacles and should be avoided as part of the function provisioning quality checks.

\section{Conclusions and Future Work}

\paragraph{Answers to research questions.}

This study evidently represents a snapshot of ongoing evolution. Summarising the details results presentation, the answers to the research
questions are as follows.

\begin{enumerate}
\item[$A_1$] (How are cloud functions offered?)
According to the principle of discoverability of (micro)services, cloud functions should be uniformly described. In practice, the pieces
of this description are scattered across different systems, making discoverability harder than necessary and partially dependent on heuristics.
Nevertheless, many cloud functions are subscribed with several metrics which contribute to the decision making process when building
a function-based serverless application.
\item[$A_2$] (How are cloud functions implemented?)
Despite attempt to support polyglot software development, cloud functions programming presents effectively a binary choice between two programming
languages, JavaScript and Python. The function binding is another binary choice. Functions are either unbound, or bound to a structural composition
of BaaS. Finally, although most function implementations have a clear representation in version control systems, the details of this representation
varies between direct file access, archived (built) file archives, and remote files.
\item[$A_3$] (Which change patterns can be recognised?)
After several years of commercial success and developer attraction around FaaS and other serverless computing offerings, a significant share of automation tasks and application functionality is covered by cloud functions. On repositories, the available cloud functions saturate the current demand, as the growth has slowed down (to around 1\% per month) but deployments are still growing at linear rate (around 7\% per month).
\end{enumerate}

\paragraph{Contributions.}

This work complements earlier mixed-method empirical studies on industrial practice in cloud function development \cite{faasproductionjournal}.
It adds a ground truth perspective with varying degrees of representativeness with most metrics related to a few hundred serverless
applications or cloud functions. Through available automated assessment scripts, assuming a further growth of the serverless computing and applications
ecosystem, the representativeness can be reevaluated at a later point in time with litte effort.
The main practical outcome of this work is a reusable evolving dataset containing metadata and code metrics from AWS SAR. Further data
has been produced as side product of the work, including code dumps and additional repository mining scripts, and will be properly
packaged and made available in the future.

\paragraph{Future work.}

This work enables future quantitative research and innovation in the following directions.

\begin{itemize}
\item Understanding microservices in practice. Cloud functions are one of the few microservice technologies which inherently
offer a service interface, rather than being a mix of services and application support
components, securing their position among ongoing microservice research \cite{DBLP:journals/corr/abs-1904-03027,DBLP:conf/asplos/GanZCSRKBHRJHPH19}.
The metrics from AWS SAR give insights into practical design and implementation decisions by developers
and could further be exploited to understand trends in microservice development including programming models and code structures.
\item Software design patterns. The resource compositions can be investigated deeper to identify not just statistical clusters,
but actual patterns of how FaaS and BaaS resources are interacting and how data flows between them. Such an analysis will have to consider
environment variables and other explicitly expressed links between the resources with intrinsic knowledge of the cloud provider's
resource model.
\item Runtime assessment of cloud functions. Moving beyond the code and configuration analysis, a generic runtime
framework for testing and performance evaluation based on AWS Lambda or with FaaS/BaaS emulation (e.g. SAM-Local and Localstack) could be constructed.
This would allow deeper insight into the execution behaviour over time, including performance deviations with varying
configurations such as FaaS memory allocation or storage backend latency.
\item Open serverless marketplaces. The conjunction of open source FaaS marketplace frameworks \cite{faasecosystems} and the open source
implementations underlying most of the functions offered on AWS SAR, compressed as single dataset by this work, allows for a rapid
prototyping of serverless marketplaces with dozens to hundreds of working functions. The openness is impeded by the reliance on
the SAM-CloudFormation mapping; through improved orchestration portability \cite{DBLP:conf/mipro/MarkoskaCRG15} and broader availability
of open source BaaS alternatives, the impediment could be overcome.
\end{itemize}

\section*{Acknowledgements}

The author thanks the student group \textit{aweSoME} (Clara-Maria Barth, Timucin Besken, Alphonse Mariyagnanaseelan, Anna Katharina Fitze, Jürg Bargetze) at University of Zurich for extending the early analysis scripts with a code contribution of GitHub metadata retrieval and several idea contributions around code analysis.

\bibliographystyle{unsrt}
\bibliography{sar-analysis}

\end{document}